\documentclass[aps,pra,twocolumn,superscriptaddress,showpacs]{revtex4}
\pdfoutput=1
\usepackage{graphicx}
\usepackage{natbib}
\usepackage[breaklinks]{hyperref}
\usepackage{amsmath}
\usepackage{mathtools}

\newcommand{\hflev}[4]{\textit{#1}$_{#2/#3}$, \textit{F}=#4}
\begin{document}


\title{Single atoms  coupled to a near-concentric cavity}
\author{Chi Huan Nguyen}
\affiliation{Centre for Quantum Technologies, 3 Science Drive 2, Singapore 117543}
\author{Adrian Nugraha Utama}
\affiliation{Centre for Quantum Technologies, 3 Science Drive 2, Singapore 117543}
\author{Nick Lewty}
\affiliation{Centre for Quantum Technologies, 3 Science Drive 2, Singapore 117543}
\author{Kadir Durak}
\affiliation{Centre for Quantum Technologies, 3 Science Drive 2, Singapore 117543}
\author{Gleb Maslennikov}
\affiliation{Centre for Quantum Technologies, 3 Science Drive 2, Singapore 117543}
\author{Stanislav Straupe}
\affiliation{Centre for Quantum Technologies, 3 Science Drive 2, Singapore 117543}
\affiliation{Faculty of Physics, M.V. Lomonosov Moscow State University, Moscow, Russia}
\author{Matthias Steiner}
\affiliation{Centre for Quantum Technologies, 3 Science Drive 2, Singapore 117543}
\affiliation{Department of Physics, National University of Singapore, 2 Science Drive 3, Singapore 117542}
\author{Christian Kurtsiefer}
\affiliation{Centre for Quantum Technologies, 3 Science Drive 2, Singapore 117543}
\affiliation{Department of Physics, National University of Singapore, 2 Science Drive 3, Singapore 117542}
\email[]{christian.kurtsiefer@gmail.com}
\date{\today}

\begin{abstract}  
Concentric cavities can lead to strong photon-atom coupling without a need for high finesse or small physical-cavity volume.
In a proof-of-principle experiment of this concept we demonstrate coupling of single Rb atoms to a 11\,mm long near-concentric cavity with a finesse $F=138(2)$. 
Operating the cavity $1.65(1)\,\mu$m shorter than the critical length, we observe an atom-cavity coupling constant~$g_0=2\pi \times 5.0(2)\,$MHz which exceeds the natural dipole decay rate~$\gamma$ by a factor $g_0/\gamma=1.7(1)$.
\end{abstract}


\maketitle
\textit{Introduction.} 
Optical cavities are widely used in a range of modern
technologies~(e.g. lasers and optical clocks) and are essential for mediating
interaction of light with other physical systems in many quantum technologies.   
In particular, by coupling atoms (or other quantum emitter) resonantly to a cavity, strongly interacting hybrid systems of light and matter can be realized~\cite{Reiserer2015}. 
This enhanced light-matter interaction is applied in quantum networks~\cite{Kimble2008,Ritter2012} and quantum metrology~\cite{Bohnet2014,Hosten2016}.  

In cavity quantum electrodynamics~(cavity QED) the conventional wisdom to realize a strongly coupled atom-cavity system employs short cavities with high finesse. 
The small mode volume~$V$ of these cavities results in a large coupling~$g_0\propto 1/\sqrt{V}$ between a single atom and a single cavity photon. 
In this situation~$g_0$ exceeds the cavity field decay rate~$\kappa$ and the dipole decay rate of the atom $\gamma$, and the light-atom interaction is dominated by the coupling to the cavity mode. 
Unfortunately, these systems are experimentally demanding due to the need of ultra-high-reflectivity coatings and sophisticated techniques to trap single atoms in these short cavities. 
However, the notion that short cavities with high finesse are inevitable has been challenged by efforts to use a particular cavity geometry, a (near-)concentric cavity, to implement cavity QED with long cavities of low finesse~\cite{Morin1994,Daul2005,Haase2006,Russo2009,Dubin2010,Aljunid2011,Chen2014,Durak2014}.  
A cavity is concentric when the separation of the two mirrors~$l_\textrm{cav}$ matches twice the radius of curvature of the mirrors~$R_C$. 
The field of the fundamental mode is tightly focused in the center of the
cavity,  leading to a small effective mode volume~$V$ while the physical size of the cavity is large~\cite{Durak2014}. 
In addition, the cavity decay rate~$\kappa\propto1/l_{cav}$ is reduced by the increased length of the cavity, which significantly eases the requirements for the mirror coatings. 
The resulting large coupling~$g_0$ and low cavity decay rate~$\kappa$ make strong coupling between single atoms and single photons feasible even with low finesse cavities. 

A second intriguing aspect of concentric cavities is that the frequencies of the higher-order transversal modes become degenerate. 
This could allow the realization of multimode cavity QED in the strong coupling regime~\cite{Wickenbrock2013}. 
Different cavity modes could then effectively interact via a commonly coupled atom -- constituting a novel platform for quantum information processing~\cite{Terraciano2009}. 
In this work we experimentally implement the idea of concentric cavity QED by trapping single $^{87}$Rb atoms in a 11\,mm long near-concentric cavity.

\begin{figure}
\centering
  \includegraphics[width=\columnwidth]{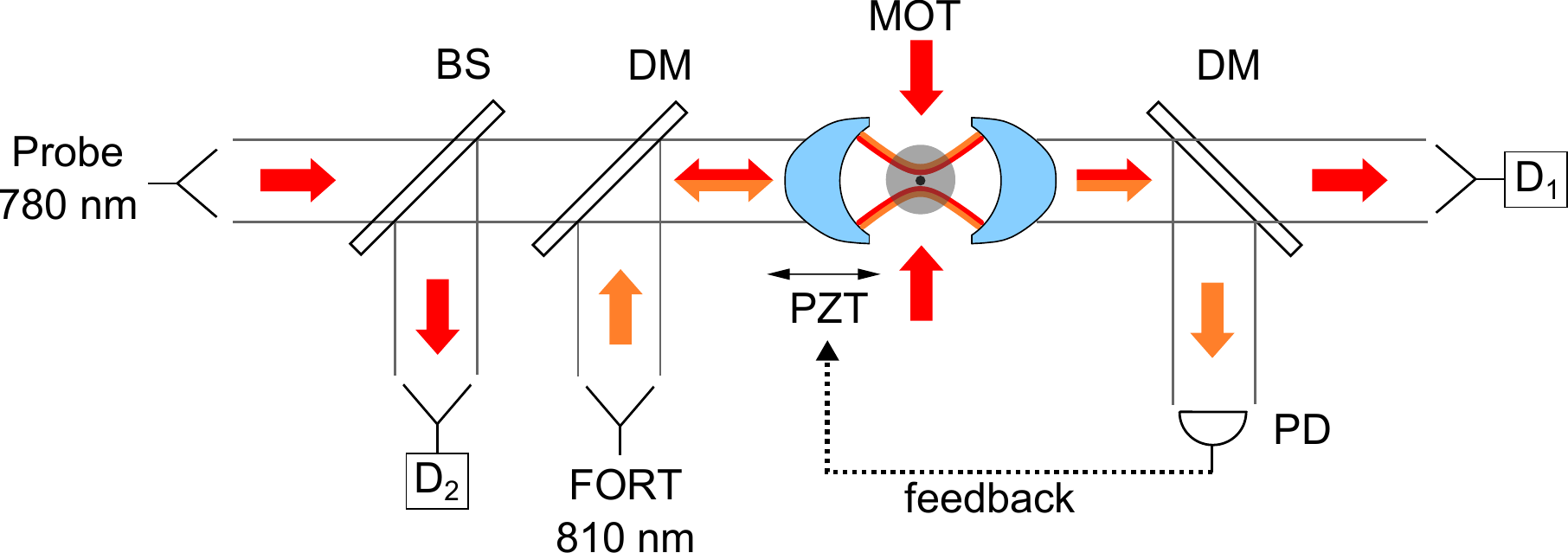}
  \caption{\label{fig:figure1} 
  Optical setup. A near resonant probe field at 780\,nm impinges on the cavity to characterize the light-atom interaction. 
  The transmitted and the reflected light is coupled into single mode fibers connected to avalanche photodetectors. 
  The cavity length is stabilized close to the concentric length by a Pound-Drever-Hall lock to a frequency stabilized 810\,nm laser. 
  The intra-cavity field at 810\,nm provides also a far-off-resonant standing-wave dipole trap for the atoms. 
  BS:~beam splitter with 70\% reflectivity, DM: dichroic mirror, PZT: 3D-piezo actuator stack, PD: photodiode, MOT: magneto-optical trap, $D_{1(2)}$: avalanche photodetectors.
}
\end{figure}

\textit{Cavity geometry.} 
The cavity is composed of two identical mirrors with a radius of curvature~$R_C=5.5\,$mm.  
To form a stable optical cavity, the stability parameter 
\begin{equation}
 g= 1 - l_\textrm{cav}/R_C
\end{equation}
needs to satisfy $0 \leq g^2 \leq 1$~\cite{Saleh2001}. Thus, 
a concentric cavity with $l_\textrm{cav} =2R_C$ is only marginally stable, and
highly susceptible to misalignment. 
However, we show that in practice the cavity can still be reliably operated extremely close to the concentric length.

We employ two methods to accurately determine the cavity
length~$l_\textrm{cav}$, which is stabilized by a Pound-Drever-Hall lock to a frequency-stabilized laser at a wavelength of 810\,nm~(Fig.~\ref{fig:figure1})~\cite{Drever1983}. 
First, we analyze the frequency spacing of the transverse cavity modes by
tuning the frequency of a probe field with a wavelength around 780\,nm. 
We find a frequency spacing $\Delta \nu_\textrm{trans}= 109(2)$\,MHz between the fundamental and first adjacent transverse mode. 
For a near-concentric cavity $\Delta \nu_\textrm{trans}$ is related to the cavity length via
\begin{equation}
\Delta \nu_\textrm{trans} = \frac{c}{2l_\textrm{cav}} \left(1- \frac{\cos^{-1}g}{\pi}  \right),
\end{equation}
where $c$ is the speed of light~\cite{Saleh2001}.
The measured mode spacing indicates a cavity length~$l_\textrm{cav}=2R_C - 1.7(1)\,\mu$m, and a cavity parameter~$g=-0.99969(2)$. 
In addition, we also use the resonance frequencies~$\nu_\textrm{780nm}$ and $\nu_\textrm{810nm}$, of two simultaneously resonant light fields to independently determine the cavity length
\begin{equation}
l_\textrm{cav}= \frac{c \Delta n}{2 (\nu_\textrm{780nm}- \nu_\textrm{810nm})} \,,
\end{equation}
where $\Delta n$ is the difference in longitudinal modes between the fields. 
For $\Delta n=1043$ we obtain $l_\textrm{cav}=2R_C-1.65(1)\mu$m and $g=-0.999700(2)$, in good agreement with the length determined from the transversal mode spacing. 
At this length, the beam waist of cavity mode is expected to be
$w_0=\sqrt{\lambda l_\textrm{cav}/(2\pi)} \left[ (1+g)/(1-g) \right]^{1/4}
=4.1\,\mu$m~\cite{Saleh2001}. 
\textit{Cavity finesse and losses.} 
We further characterize the cavity by the transmission and reflection of the 780\,nm probe field~(Fig.~\ref{fig:figure1}). 
To achieve good mode matching between the fundamental mode of the cavity and
the external probe field with Gaussian profile, we implement a so-called
anaclastic lens design~\cite{ibnisahl, Rashed:1990}: 
The non-reflective back end of the mirrors have an ellipsoidal shape to act as an aspheric surface, converting the plane wavefront of a collimated Gaussian input beam to a converging spherical wavefront~\cite{Durak2014}. 
\begin{figure}
\centering
  \includegraphics[width=\columnwidth]{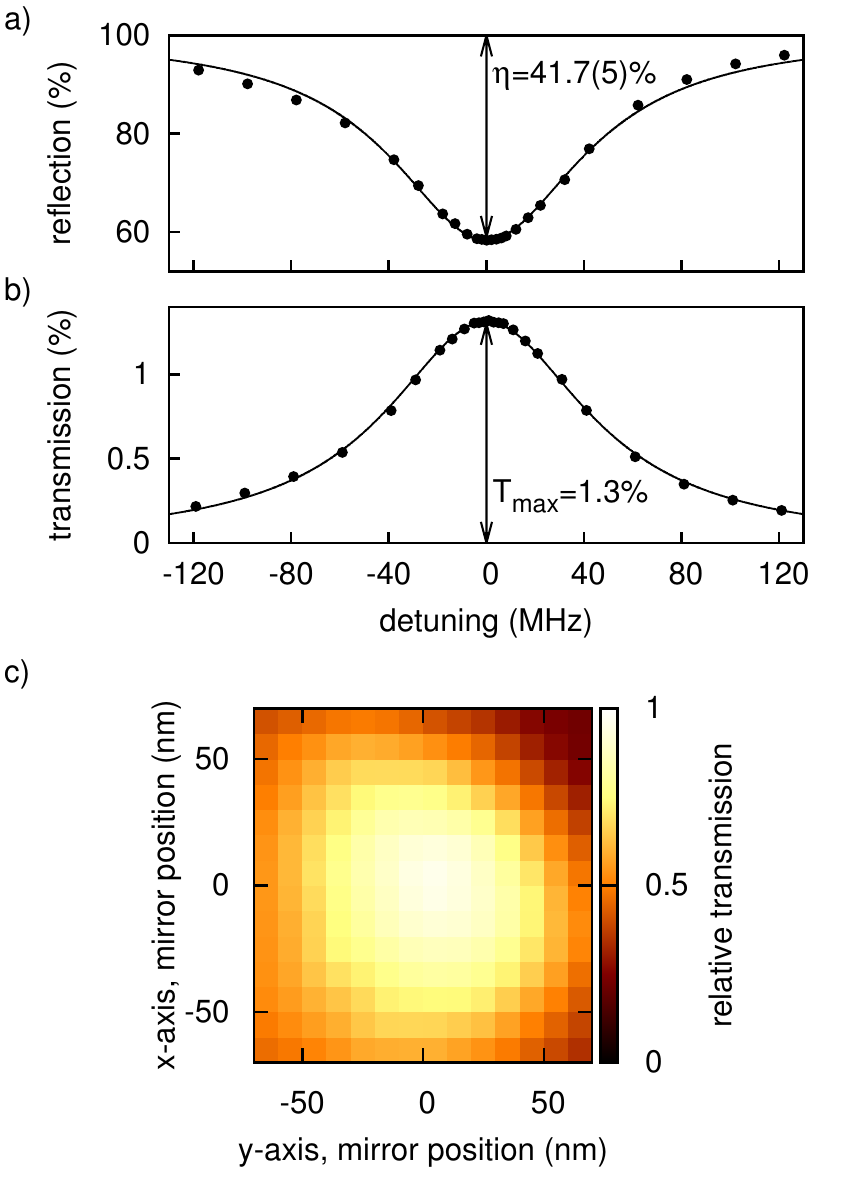}
  \caption{\label{fig:figure2} 
  a) Reflection and b)~transmission~ of a 780\,nm probe field measured after
  mode cleaning with the single-mode fiber. 
  Solid lines are Lorentzian fits. 
  c) Normalized cavity transmission as one mirror is transversally
  displaced. Throughout the experiment, the cavity length is actively stabilized to be resonant with the probe field. 
}
\end{figure}

Tuning the probe frequency, we record the reflection- and transmission
spectrum, which we fit to Lorentzian profiles. We obtain a full-width-at-half-maximum~(FWHM) of $95(3)\,$MHz and $99(1)\,$MHz, respectively~(Fig.~\ref{fig:figure2}a-b). 
Conservatively, we attribute the transmission linewidth to the fundamental mode of the cavity,~$2\kappa=2\pi\times99(1)\,$MHz, corresponding to a cavity finesse~$F=\pi c/(2\kappa l_\textrm{cav})=138(2)$~\cite{Saleh2001}.  
Originally, the finesse of the mirrors was higher $F\geq 500$, but dropped after bake-out of the vacuum chamber and operating the Rubidium dispenser.  
From the finesse and the nominal transmission~$T=0.5\%$  of the mirrors, we
deduce a round-trip absorption loss~$L$, the maximum in-coupling
efficiency~$\eta$, and resonant transmission~$T_\textrm{max}$ in the usual
way~\cite{Hood2001} via
\begin{eqnarray}
L=2\pi/F-2T=3.6(1)\%,\label{eq:finesse}
\\
\eta=1-L^2/(2T+L)^2=39(1)\%,\label{eq:eta}
\\
T_\textrm{max}= 4 T^2/(2T+L)=4.7(2)\%\label{eq:Tmax}.
\end{eqnarray}
In a direct measurement, we observe a cavity incoupling efficiency of~$\eta=41.7(5)\%$, which agrees with Eq.~\ref{eq:eta} and demonstrates that the anaclastic design provides excellent mode matching between the probe field and the fundamental cavity mode~(Fig.~\ref{fig:figure2}a).  
The resonant transmission~$T_\textrm{max}=4.6(2)\%$, measured directly after the cavity, is also in good agreement with Eq.~\ref{eq:Tmax}. 
The transmission shown in Fig.~\ref{fig:figure2}b is lower because the transmitted light is coupled into a single mode fiber before detection.

\textit{Cavity stability.} 
Approaching the concentric length~$l_\textrm{cav} \rightarrow 2R_C $, the
cavity becomes only marginally stable, and consequently is highly sensitive to small misalignments. 
Therefore, one of cavity mirrors is placed on a 3D-piezo actuator stack which allows us to move the mirror $5\mu$m in each direction.  
Figure~\ref{fig:figure2}c shows the resonant transmission of the 780\,nm probe
field as we tune the transversal position of one mirror;
the transmission shows a FWHM of 59(3)\,nm along both transverse directions. 
This high sensitivity to the transversal alignment requires active
stabilization to compensate drifts caused, for example, by temperature fluctuations. 
Using the transmission of the 780\,nm and 810\,nm light as feedback signals, we optimize the transversal mirror position every 15\,minutes. 

\begin{figure}
\centering
  \includegraphics[width=\columnwidth]{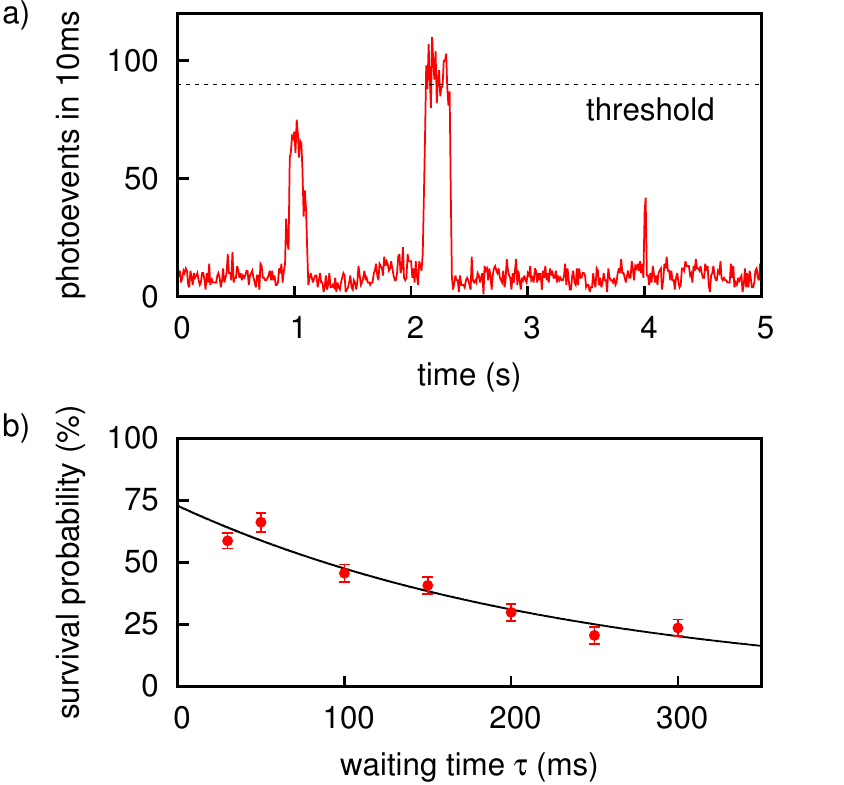}
  \caption{\label{fig:figure3} 
a) Typical trace of detection events at detector~$D_1$ with an atomic cloud in the MOT inside the cavity. 
The cooling light is 10\,MHz red-detuned from the natural 5\hflev{S}{1}{2}{2} to 5\hflev{P}{3}{2}{3} transition frequency. 
The sudden increase of fluorescence indicates the entering of an atom into the FORT.  
b)~Lifetime of single atoms in FORT without cooling light for a time $\tau$. The
solid line represents an exponential fit with a $1/e$-lifetime
$t_0=230(30)\,$ms.
}
\end{figure}

\textit{Determining the atom-cavity interaction.}
To probe the light-atom interaction, we prepare a cold ensemble of $^{87}$Rb atoms in a  magneto-optical trap~(MOT). 
The large physical separation of the two mirrors allows us to form the MOT inside the cavity. 
Atoms from the MOT are probabilistically loaded into the far off-resonant dipole trap~(FORT) created by the intra-cavity field of the 810\,nm light used to stabilize the cavity length. 
To account for the light shift induced by the FORT, the cavity length is set so that the resonance frequency is 22\,MHz higher than the 5\hflev{S}{1}{2}{2} to 5\hflev{P}{3}{2}{3} transition. 
While operating the MOT, we detect the coupling of individual atoms to the fundamental cavity mode by the sudden increase of fluorescence at detector~$D_1$~\cite{Mabuchi1996,Hood1998,Ye1999}.    
Figure~\ref{fig:figure3} shows a typical fluorescent trace during the loading process, exhibiting a telegraph signal characteristic for single atom loading. 
From the low frequency of loading events we infer that the probability of simultaneously loading two atoms in the center region of the cavity to be negligible. 
The lifetime of an atom in the trap is determined by switching off the cooling
beams after a loading event for different waiting times~$\tau$. The
survival probability $p(\tau)$ decays exponentially with a characteristic
$1/e$ lifetime of $230(30)\,$ms determined from a fit~(Fig.~\ref{fig:figure3}b). 

The single atom--cavity coupling~$g_0$ can be determined from the cavity transmission and reflection~\cite{Raizen1989,Thompson1992}. 
For a weak coherent beam, the coefficients for intensity transmission~$T(\omega)$ and reflection~$R(\omega)$ are given by
\begin{eqnarray}
T(\omega) =   \left| \frac{\kappa_T \left( i \Delta_a + \gamma\right)}{\left( i \Delta_c + \kappa\right)\left( i \Delta_a + \gamma\right)+g^2_0} \right|^2,~~ \label{eq:T}
\\
R(\omega) =  \left| 1 - \frac{2\kappa_T \left( i \Delta_a + \gamma\right)}{\left( i \Delta_c + \kappa\right)\left( i \Delta_a + \gamma\right)+g^2_0} \right|^2,~~ \label{eq:R}
\end{eqnarray}
with a cavity field decay rate through each mirror~$\kappa_T= T \pi c/ l_\textrm{cav}$, and the detuning $\Delta_{c,(a)} = \omega-\omega_{c,(a)}$ of the driving laser with respect to the cavity (atomic transition) frequency $\omega_{c,(a)}$, respectively~\cite{Reiserer2015}.
Once an atom is loaded, we use an experimental sequence that alternates between 1\,ms of probing the cavity transmission, and 1\,ms of laser cooling by the MOT beams. 
The detected photoevents during the cooling cycle are used to check whether the atom is still present.  

\begin{figure}
\centering
  \includegraphics[width=\columnwidth]{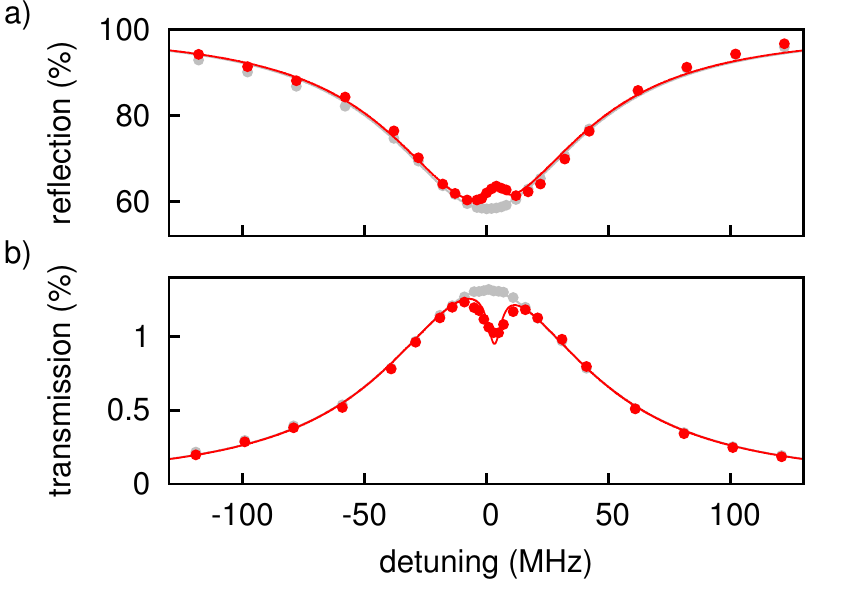}
  \caption{\label{fig:figure5} 
  Onset of the normal-mode splitting in the a)~reflection and b)~transmission spectrum when an atom is trapped in the FORT. 
  Red solid lines are fits based on Eq.~\ref{eq:T}. 
  For comparison the empty cavity reflection/transmission~(Fig.~\ref{fig:figure2}a) is shown in gray.
}
\end{figure}

The atom-light interaction is revealed in the reflection-- and transmission
spectrum obtained by tuning the frequency of the probe laser. 
When an atom is present, the spectra show the onset of the normal-mode splitting~(Fig.~\ref{fig:figure5}, red circles). 
From a least-squares fit of the transmission spectrum to Eq.~\ref{eq:T} with two free parameters, we obtain an interaction strength~$g_0=2\pi \times5.0(2)\,$MHz and a frequency offset~$\omega_\textrm{off}=\omega_{c}-\omega_{a}=2\pi\times 3.4(3)\,$MHz between cavity and atomic resonance. 
The amplitude of the fit function~$T(\omega)$ is set to the
independently determined maximum transmission of the empty cavity.
From $g_0$, the cavity linewidth~$2\kappa=2\pi \times 99(1)$\,MHz, and the natural transition linewidth~$2\gamma=2\pi\times 6.07$\,MHz, we obtain the single atom cooperativity $C_0=g^2_0/(2\kappa\gamma)=0.084(4)$. 

The reflection spectrum is analyzed in a similar way by fitting to Eq.~\ref{eq:R}.   
For this, we use three fit parameters, $g_0=2\pi \times4.6(4)\,$MHz, the
frequency offset~$\omega_\textrm{off}=2\pi \times 4.4(7)\,$MHz, and the
reflected power far away from the atom/cavity resonance. 
The fits of Eq.~\ref{eq:T}-\ref{eq:R} to the transmission and reflection
reproduce the observed values very well (Fig.~\ref{fig:figure5}, solid lines),
and lead to similar values for the atom-cavity coupling constant~$g_0$ and the frequency offset~$\omega_\textrm{off}$. 

The experimentally obtained value for $g_0$ is lower than expected for a clean
two-level atom from the cavity geometry~$g_0= \sqrt{3 \lambda^2 c \gamma/(4\pi V)} = 2\pi\times12.1\,$MHz~\cite{Reiserer2015}. 
We attribute this partly to the fact that in this experiment, the atom is
prepared by the MOT beams in a random spin state~$m_F$ of the
5\hflev{S}{1}{2}{2} manifold before the transmission is probed with a linearly
polarized probe field.
Averaging over the corresponding Clebsch-Gordan coefficients, we estimate that
the atom-cavity coupling should be a factor $\sqrt{2}$ larger for a circularly
polarized probe field driving an atom prepared in the 5\hflev{S}{1}{2}{2},
$m_F$=2 on a transition to the  5\hflev{P}{3}{2}{3}, $m_F$=3 state.

\textit{Discussion and conclusion.} 
Our experiment demonstrates the prospects and challenges of concentric cavity QED. 
The realization of atom-cavity coupling exceeding the natural dipole decay
rate by a factor $g_0/\gamma=1.7(1)$ could stimulate further efforts employing concentric cavities.  
The coupling observed in this proof-of-principle experiment is already similar
to many state-of-the-art experiments in the strong coupling regime, but with a
 cavity two orders of magnitude shorter~\cite{Reiserer2015}. Only in in very short fiber cavities, significantly larger values of $g_0/\gamma$ have been demonstrated~\cite{Gehr2010}.
Going closer to the concentric length~$l_\textrm{cav}\rightarrow2R_C$ should
increase the interaction strength even further. 
We estimate that a ratio $g_0/\gamma\geq4$ can be achieved for
$l_\textrm{cav}\approx2R_C- 100$\,nm. When stabilizing the cavity near this
point, we currently observe that the cavity finesse and transmission drop,
possibly due to deviations of the mirror from an ideal spherical surface, and
stronger coupling of the probe field to other higher-order transversal cavity
modes. 

Even without operating closer to the concentric length, we expect that a single atom cooperativity above unity can be reached by modestly increasing the finesse to $F=1000$ and performing the probing on a cyclic transition. 
A medium cavity finesse of $F\geq4500$ would put this system into the single atom-single photon strong coupling regime. 
While our experiments are performed with single neutral atoms, concentric
cavities are also interesting for other quantum systems:
examples are trapped ions~\cite{Tauschinsky2010} and Rydberg
atoms\cite{Abel2011,Brownnutt2015}, which both are
experimentally difficult to hold within short cavities due to the electric field noise near dielectric mirrors. 

\begin{acknowledgments}This work was supported by the Ministry
  of Education in Singapore (AcRF Tier 1) and the National Research
  Foundation, Prime Minister’s office (partly under grant no
  NRF-CRP12-2013-03). M.S. acknowledges support by the Lee Kuan Yew
  Postdoctoral Fellowship. 
\end{acknowledgments}
\bibliographystyle{apsrev4-1}
%

\end{document}